\documentclass[twocolumn, prx, superscriptaddress,notitlepage]{revtex4-2}
\usepackage{graphicx}
\usepackage{dcolumn}
\usepackage{bm}
\usepackage[usenames,dvipsnames]{color}
\usepackage[most]{tcolorbox}
\usepackage{multirow}
\usepackage{gensymb}
\usepackage[normalem]{ulem}
\usepackage{CJK}
\usepackage{comment}
\usepackage[colorlinks, linkcolor=blue,anchorcolor=blue,citecolor=blue,urlcolor=blue]{hyperref}
\usepackage{amssymb}
\usepackage{pifont}
\usepackage{physics}
\usepackage{natbib}

\usepackage{booktabs}
\usepackage{colortbl}
\newcommand{\xfactor}{\alpha} 
\newcommand{\ham}{{\mathcal{H}}}
\newcommand{\sx}{\sigma_x}
\newcommand{\sy}{\sigma_y}
\newcommand{\sz}{\sigma_z}

\usepackage{color,soul}
\usepackage[mathscr]{euscript}
\begin{document}
\begin{CJK*}{UTF8}{}
\title{
Individual-atom control in array through phase modulation}

\author{Guoqing Wang}\email[]{gq\_wang@mit.edu}
\thanks{These authors contributed equally.}
\affiliation{
   MIT-Harvard Center for Ultracold Atoms and Research Laboratory of Electronics, Massachusetts Institute of Technology, Cambridge, MA 02139, USA}
\affiliation{
   Department of Nuclear Science and Engineering, Massachusetts Institute of Technology, Cambridge, MA 02139, USA}
\affiliation{
   Department of Physics, Massachusetts Institute of Technology, Cambridge, MA 02139, USA}
\author{Wenchao Xu}
\thanks{These authors contributed equally.}
\affiliation{
   Institute of Quantum Electronics, ETH Z\"{u}rich, Z\"{u}rich 8093, Switzerland}
 \affiliation{
   Paul Scherrer Institut, CH-5232 Villigen PSI, Switzerland}  
\author{Changhao Li}
\affiliation{
   MIT-Harvard Center for Ultracold Atoms and Research Laboratory of Electronics, Massachusetts Institute of Technology, Cambridge, MA 02139, USA}
\affiliation{
   Department of Nuclear Science and Engineering, Massachusetts Institute of Technology, Cambridge, MA 02139, USA}

\author{Vladan Vuleti\'c}\email[]{vuletic@mit.edu}
\affiliation{
   MIT-Harvard Center for Ultracold Atoms and Research Laboratory of Electronics, Massachusetts Institute of Technology, Cambridge, MA 02139, USA}
\affiliation{
   Department of Physics, Massachusetts Institute of Technology, Cambridge, MA 02139, USA}

\author{Paola Cappellaro}\email[]{pcappell@mit.edu}   
\affiliation{
   MIT-Harvard Center for Ultracold Atoms and Research Laboratory of Electronics, Massachusetts Institute of Technology, Cambridge, MA 02139, USA}
\affiliation{
   Department of Nuclear Science and Engineering, Massachusetts Institute of Technology, Cambridge, MA 02139, USA}
\affiliation{
   Department of Physics, Massachusetts Institute of Technology, Cambridge, MA 02139, USA}
 
\begin{abstract}
Performing parallel gate operations while retaining low crosstalk is an essential step in transforming neutral atom arrays into powerful quantum computers and simulators. Tightly focusing control beams in small areas for crosstalk suppression is typically challenging and can lead to imperfect polarization for certain transitions. We tackle such a problem by introducing a  method to engineer single qubit gates through phase-modulated continuous driving. Distinct qubits can be individually addressed to high accuracy by simply tuning the modulation parameters, which significantly suppresses crosstalk effects. When arranged in a lattice structure,  individual control with optimal crosstalk suppression is achieved. With the assistance of additional addressing light or multiple modulation frequencies, we develop two efficient implementations of parallel-gate operations.  Our results pave the way to scaling up atom-array platforms with low-error parallel-gate operations, without requiring complicated wavefront design or high-power laser beams.
\end{abstract}

\maketitle
\end{CJK*}	

\section{Introduction} Arrays of individual atoms trapped in optical tweezers have emerged as an attractive architecture for implementing quantum computation and simulation~\cite{morgado_quantum_2021,saffman_quantum_2010}. Current state-of-the-art atom array platforms have achieved {global single-qubit operation with} 99.99\% fidelity and two-qubit gate operation with 99.5\% fidelities~\cite{sheng_high-fidelity_2018,levine_parallel_2019,levine_dispersive_2022,evered_high-fidelity_2023,ma_high-fidelity_2023,scholl_erasure_2023}. Still, performing scalable, arbitrary local gate operations over a subset of atoms inside an atom array {without crosstalk} is an outstanding challenge. The generation of arrays of rapidly switchable and reconfigurable laser beams that can address atoms individually is a demanding task for optical modulators, especially in view of the requirements on the intensity and frequency stability of these laser beams.

{Currently, there are two major strategies for performing local qubit operations. The first one relies on a separated manipulation region into which the target atoms are transported, and that is addressed by global beams~\cite{bluvstein_quantum_2022}. The transport time of this strategy unavoidably scales up with the system size. The second approach  introduces local light shifts with individual off-resonant addressing beams, such that only a subset of atoms inside the array are resonant with the global addressing beam~\cite{omran2019generation,labuhn2014single,de2017optical,burgers2022controlling}. To reduce crosstalk and to suppress the error rate due to light scattering, a large detuning and high power are preferred, which is technically challenging.}

\begin{figure}[htbp]
\includegraphics[width=0.45\textwidth]{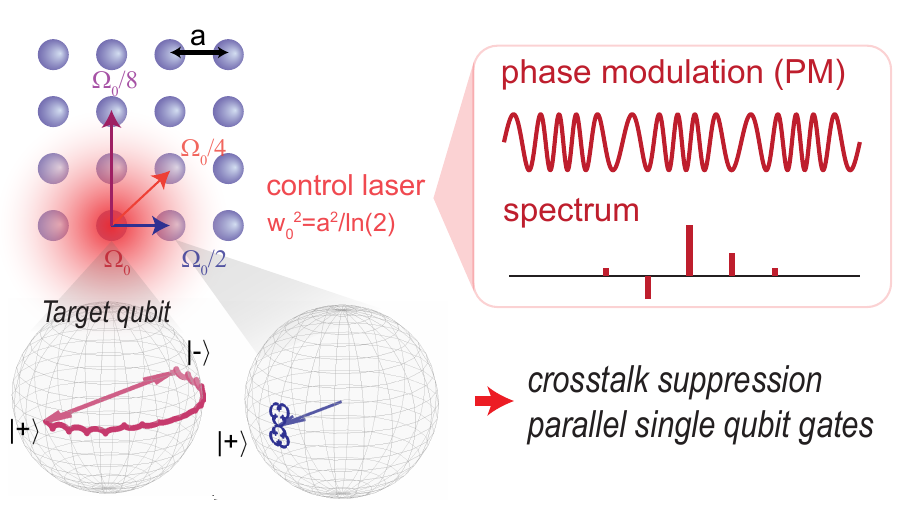}
\caption{\label{Fig0} \textbf{Low-crosstalk gates in atom arrays.} By modulating the phase of a loosely focused local addressing beam, a rotational gate is applied to the target atom while deeply suppressing the crosstalk on nearby atoms. The evolution trajectories on Bloch spheres showcase a crosstalk-suppressed $Z$ gate for the target and nearby atoms. }
\end{figure}

In this work, we propose a novel scheme to implement parallel single-qubit gate operations on the target qubits while suppressing crosstalk effects due to power leakage on the rest of the qubits. By engineering a phase-modulated concatenated continuous driving, the desired quantum gate is selectively applied to the target qubit, while the rest experience identity operations. As shown in Fig.~\ref{Fig0}, for atom arrays arranged in a lattice structure, even better crosstalk suppression can be achieved by properly choosing the spacing between optical tweezers. Based on such a technique, we further propose two new schemes to achieve parallel gate operations on a subset of atoms by either introducing additional light shifts or tuning multiple modulation frequencies. In comparison to existing AC stark shift-based methods, our scheme does not require a far-detuned high-power laser, which paves the way to achieving large-scale quantum computation with low-error parallel-gate operations.

\begin{figure}[htbp]
\includegraphics[width=0.49\textwidth]{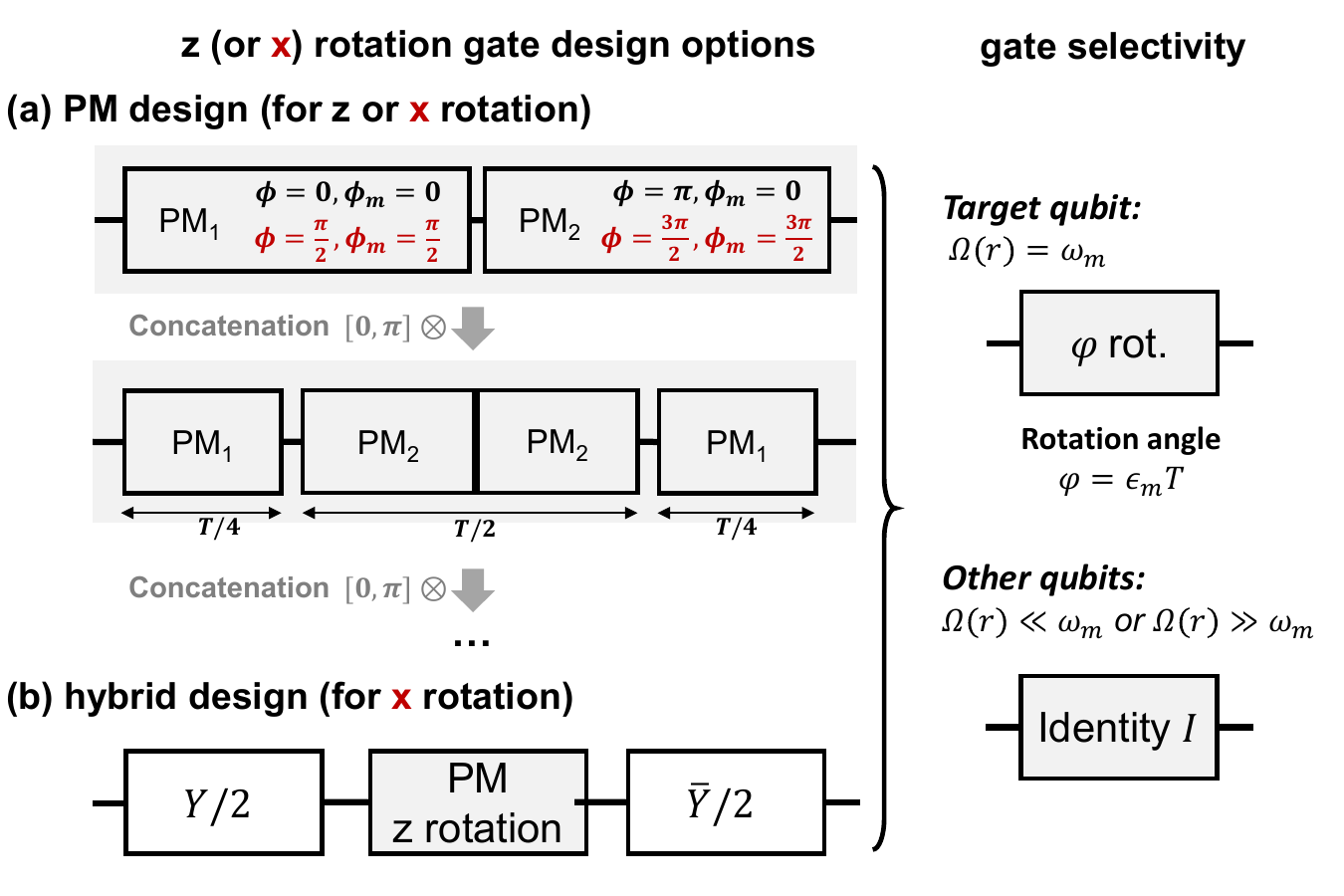}
\caption{\label{Fig1}\textbf{Single qubit rotation gate design.} Each block represents the application of the PM drive with the specified phases $\phi,\phi_m$. For rotations along $x$, a hybrid design option combines a PM $z$ rotational gate with bare $Y$ rotations.}

\end{figure}
\section{Constructing low-crosstalk gate using phase modulation}
A variety of control techniques have been exploited to engineer quantum gates in the presence of control constraints. The most-studied approach for protecting a quantum operation against these limitations are  composite pulses~\cite{freeman1998spin}, which combine several imperfect pulses to construct an operation with smaller error~\cite{brown_arbitrarily_2004,mc_hugh_sixth-order_2005,torosov_arbitrarily_2018}. Tailored to different applications, composite pulses can be designed to have broadband, narrowband, or passband features~\cite{WIMPERIS1994221,torosov_composite_2015}, and can be made robust to non-unitary errors~\cite{Khodjasteh_PhysRevA.86.042329,Khodjasteh_PhysRevLett.102.080501}.  Based on available primitive gates, a systematic and efficient methodology for composite gate design has been developed~\cite{low_methodology_2016}. Efforts have also been devoted to improving the smoothness of the temporal shape~\cite{torosov_smooth_2011,bartels_smooth_2013}.

The errors arising in the gate of operation of atom arrays are however distinct from many of the situations that have been previously addressed, and thus call for new solutions. 
The crosstalk in atom array systems is predominantly induced by the leakage of the laser power to neighboring qubits due to the finite size focus of the laser beam. Such an issue originates from a dilemma on the choice of beam waist: while a very small beam waist is preferred to reduce the crosstalk with nearby atoms, it raises challenges in maintaining position stability and polarization purity. In contrast to methods that require additional  light shifts to differentiate qubits and suppress their crosstalk, here we propose to engineer the global phase of the control beam, exploiting so-called \textit{concatenated continuous driving} (CCD), which was previously developed to protect qubit coherence~\cite{caiRobustDynamicalDecoupling2012,farfurnikExperimentalRealizationTimedependent2017,wangCoherenceProtectionDecay2020,wang_observation_2021-symmetry,wang_observation_2021} and design optimal control pulses in solid-state systems~\cite{TianQuantumOptimalControlPhysRevA.102.043707,khanejaUltraBroadbandNMR2016}.
The phase modulation (PM) applied to the control field is described by the following single-qubit Hamiltonian
\begin{equation}
    \ham_{PM}=\frac{\omega_0}{2}\sz+\Omega\cos\left(\omega t +\phi-\frac{2\epsilon_m}{\omega_m}\sin(\omega_m t+\phi_m)\right)\sx,
    \label{Hamiltonian_PhaseMod}
\end{equation}
where $\omega$ is set to the qubit frequency $\omega_0$, $\Omega=\Omega(r)$ is the spatially varying driving strength, and $\epsilon_m,\omega_m,\phi_m$ are spatially independent modulation parameters.
In the rotating frame defined by $U=\exp\left[-i\left(\frac{\omega t}{2}\sz-\epsilon_m \frac{\sin(\omega_m t+\phi_m)}{\omega_m}\sz \right)\right]$, the Hamiltonian  $\ham_I=U^\dagger \ham_{PM} U-U^\dagger i\frac{d}{dt}U$  is
\begin{equation}
        \ham_I=\frac{\Omega}{2}\sigma_x^\prime+\epsilon_m\cos(\omega_mt+\phi_m)\sz,
    \label{HI_PhaseMod}
\end{equation}
where $\sigma_x^\prime=\sx \cos\phi+ \sy \sin\phi$.

We note that this Hamiltonian can be made time-independent by transforming to a second interaction picture (rotating) frame defined  by $\omega_m\sigma_x^\prime/2$, and taking the  rotating wave approximation ($\epsilon_m\ll\omega_m$). We then obtain the Hamiltonian
\begin{equation}
    \ham_{I,2}=\frac{\Omega-\omega_m}{2}\sigma_x^\prime+\frac{\epsilon_m}{2}\sigma_z^\prime
    \label{eq:HI2}
\end{equation}
where $\sigma_z^\prime=\cos\phi_m\sz-\sin\phi_m\sy^\prime$.

Consider, e.g., the goal of applying a  rotation of angle $\varphi$ about the $z$ axis: our scheme implements this by driving a $\varphi$ rotation in the second rotating frame. We set the modulation frequency to the resonance condition $\omega_m=\Omega$ and the modulation phase $\phi_m=0$ such that $\sigma_z^\prime=\sigma_z$. By further setting the modulation amplitude to $\epsilon_m=\omega_m\varphi/(2\pi k)$ with $k$ any positive integer, a $\varphi$ rotation along $z$ is achieved for a total evolution time $T=\varphi/\epsilon_m$. Furthermore, the (second) rotating frame transformation is an identity operation as $\omega_m T=2\pi k$, thus a $\varphi$ rotation along $z$ is also applied in the first rotating frame.

The benefits of such a method is manifested in its selectivity to the resonance condition $\Omega=\omega_m$, which is equivalent to selectivity to the target qubit, thanks to the spatial dependence of $\Omega$. For the rest of the qubits, which should ideally undergo the identity operation under the applied control, the actual rotation  is along a direction close to $x^\prime$ at a rate $\sqrt{(\Omega-\omega_m)^2+\epsilon_m^2}$, since $|\Omega-\omega_m|\gg\epsilon_m$. Such an evolution can be simply canceled by alternating between $\phi=0$ (giving $\sigma_x^\prime=\sigma_x$)  and $\phi=\pi$ (giving $\sigma_x^\prime=-\sigma_x$) in the two halves of the total evolution duration, which gives an effective identity operation (Fig.~\ref{Fig1}(a)). 

Similar designs can engineer single-qubit rotation gates along other directions by tuning the modulation phases $\phi,\phi_m$. To engineer rotations along $x$, we can set the modulation phase $\phi_m=\pi/2,3\pi/2$ and $\phi=\pi/2,3\pi/2$ for the two halves of the evolution such that the total rotation is along $\sz^\prime =\sy^\prime=\sx$. An improved design achieves a better crosstalk suppression at small power leakage: this so-called \textit{hybrid design} interleaves the PM-designed $z$ rotation in between two bare $\pi/2$ rotations about $y$ or $-y$ ($Y/2$ and $\Bar{Y}/2$ gates in Fig.~\ref{Fig1}(b)).

\begin{figure}[htbp]
\includegraphics[width=0.49\textwidth]{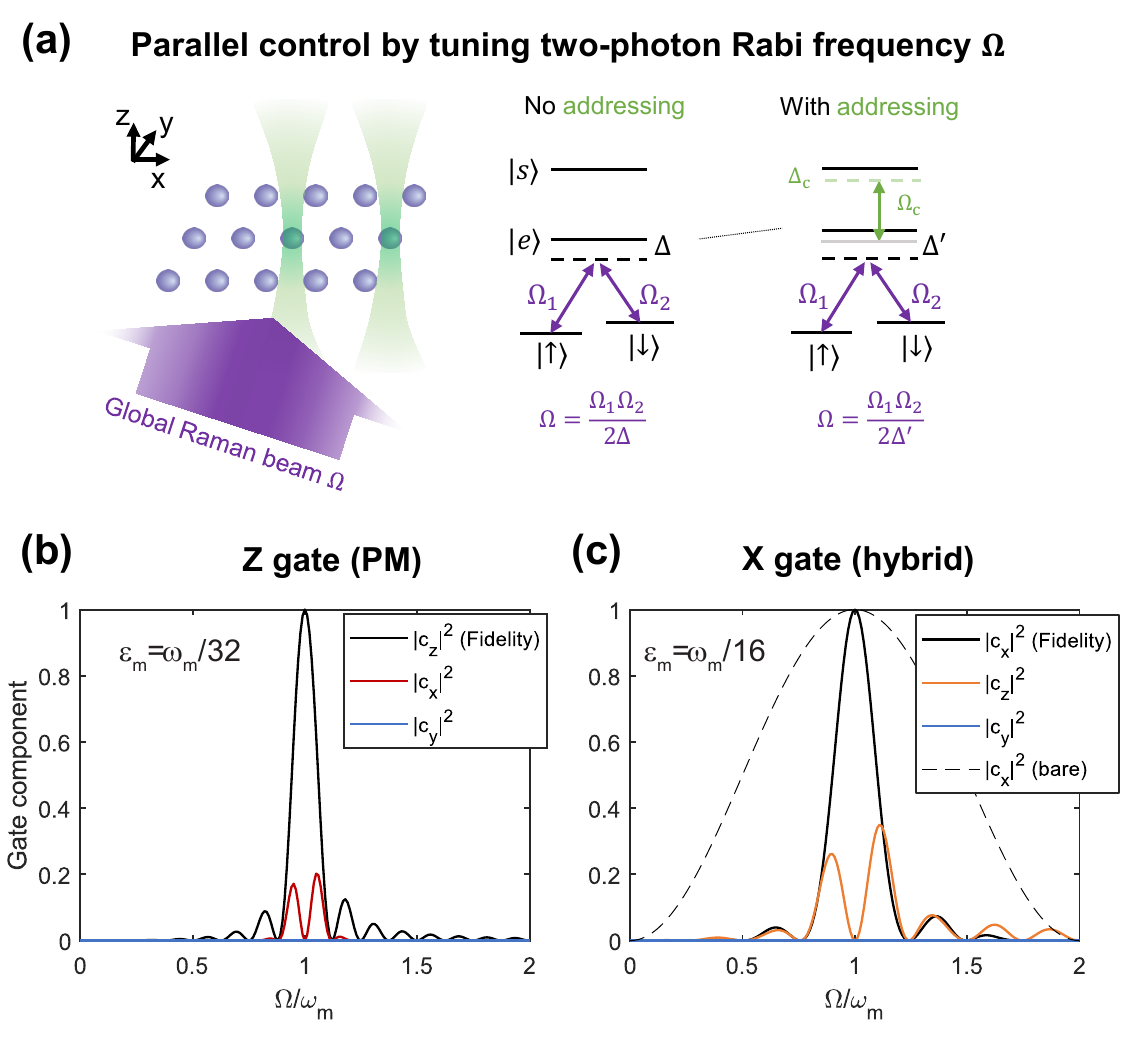}
\caption{\label{Fig2}\textbf{Parallel control of atom array with additional light shifts.} (a) Parallel single qubit control by tuning $\Omega$. Global qubit operation beams are sending towards atoms to drive a Raman transition between the two qubit states. Individual addressing beams couple the intermediate state towards another state to induce AC Stark energy shift. (b) Fidelity of the PM design $Z$ gate (PM1-PM2-PM2-PM1, same for c) as a function of Rabi frequency normalized by the modulation frequency $\Omega/\omega_m$. (c) Fidelity of the hybrid design $X$ gate as a function of Rabi frequency.}
\end{figure}

Using $Z$ (PM design) and $X$ (hybrid design) gates as examples, we numerically simulate the evolution $U=\mathcal{T}e^{-i\int H_I(t)dt}$ in the first rotating frame, where $\mathcal{T}$ is the time ordering operator. Decomposing $U$ as $U=c_0I+c_x\sx+c_y\sy+c_z\sz$, we plot the values of $|c_x|^2,|c_y|^2,|c_z|^2$ as a function of $\Omega/\omega_m$, which characterize the crosstalk effect. The values at $\Omega/\omega_m=1$ indicate the gate performance for the target qubit, which shows high-fidelity $Z$ and $X$ operations in Fig.~\ref{Fig2}(b) and (c), respectively. For the rest of the qubits with $\Omega/\omega_m\ll 1$, crosstalk effects for $X$ gate are suppressed in comparison to the bare $X$ gate without PM (dashed line).

The principle of our selective operation strategy is similar to the light shift-based method in the sense that the driving term is on resonance with the target qubit and off-resonance with the spectator qubits. The main difference is that the off-resonant term in light-shift methods induces energy shifts in a second-order perturbation manner, while the effective ``energy shift" in our method is linearly dependent on the drive amplitude through the first-order coupling. Thus, large laser power is not required in our method to generate such an energy shift. The bandwidth of the designed gate is then set by the modulation amplitude $\epsilon_m$ as shown in Fig.~\ref{Fig2}(b,c). To suppress crosstalk effects, the detuning term $\omega_m-\Omega$ for spectator qubits should be much larger than the modulation amplitude $\epsilon_m$ (see Eq.~\eqref{eq:HI2}). 

Making use of the periodic zero-crossing feature of the gate components  shown in Figs.~\ref{Fig2}(b,c), one can arrange atom arrays into special spatial patterns to achieve even better suppression performance. When the atoms are arranged in a square lattice as shown in Fig.~\ref{Fig2}(b), the crosstalk effects can be optimally suppressed by setting the radius $r_0$ of the Rabi frequency spatial profile $\Omega(r)\propto e^{-\frac{r^2}{r_0^2}}$ (applied by a Gaussian laser beam) to $a=r_0\sqrt{\ln 2}$ where $a$ is the lattice constant. Under this condition, the nearest neighbor has one half of the driving amplitude while the second nearest neighbor has a quarter in comparison to the target qubit, and gate amplitudes ($x,y,z$ rotation components) at these sites are zero, as shown in the simulation in Figs.~\ref{Fig2}(b,c), indicating optimal crosstalk suppression performance.

\section{Parallel control schemes}
\subsection{Parallel control of hyperfine qubits with additional light shifts}
Until now we considered to encode the qubit in energy levels associated with single-photon transitions (optical qubits). However,  many quantum computing platforms are based on long-lived hyperfine qubits controlled by two-photon Raman transitions. In the following, we show that our approach can be used to implement parallel single qubit operations in these platforms, with the assistance of additional individual-addressing light as shown in Fig.~\ref{Fig2}(a). 

Let us consider a $Z$ gate engineered through a two-photon process by two global laser beams, as shown in Fig.~\ref{Fig2}(a) with a (two-photon) Rabi frequency $\Omega_{eff}=\Omega_1\Omega_2/(2\Delta)$. Off-resonant addressing beams are applied on target qubits. These addressing beams mix the intermediate state with a fourth state to change the effective detuning in the two-photon transition, thus modifying the Rabi frequency of the hyperfine qubit. Combined with the selectivity of the target operations over the Rabi frequency shown in Figs.~\ref{Fig2}(b,c), the target operations are then applied to atoms with (or without) addressing beams, while identity operations are applied to the spectators. With the periodic zero-crossing spectrum feature, the additional light shifts only need to change the effective Rabi frequency by an amount on the order of $\sim\epsilon_m$ instead of $\Omega$.

More quantitatively, in the rotating frame defined by the drive frequencies, the system can be described by a 4-level Hamiltonian $H=-\Delta\ket{e}\!\bra{e}-(\Delta+\Delta_c)\ket{s}\!\bra{s}+[\Omega_1\ket{\uparrow}\bra{e}+\Omega_2\ket{\downarrow}\!\bra{e}+\Omega_c\ket{e}\!\bra{s}+h.c.]/2$ with $\ket{\downarrow},\ket{\uparrow}$ denoting the qubit states and $\ket{e},\ket{s}$ denoting the intermediate and additional states. 
When $\Delta,\Delta-\Delta_c\gg\Omega_1,\Omega_2$, we can adiabatically eliminate the two excited states $\ket{e}$ and $\ket{f}$ and simplify the Hamiltonian to the qubit subspace (see Appendix.~\ref{App:light shifts} for details).
The effective Rabi rate between the qubit states is
\begin{equation}
\Omega_\text{eff} =\frac{\Omega_1\Omega_2}{2\Delta-\frac{\Omega_c^2}{2(\Delta+\Delta_c)}}
\end{equation}
To modify the effective Rabi rate by a factor of $\xfactor$, we can choose the detuning as $\Delta_c=-\Delta+{\xfactor \Omega_c^2}/{[4\Delta(\xfactor-1)]}$.

When using acousto-optic deflectors (AOD) to address individual atoms, undesirable position-dependent frequency shifts $\delta_c$ are introduced, which lead to a Rabi frequency change with a ratio $-\frac{4\Delta (1-\xfactor)^2}{\Omega_c^2} \delta_c$. Therefore, small $\Delta$ and large $\Omega_c$ are preferred. If we consider the two hyperfine ground states of $^{171}$Yb coupled through the triplet clock state $^3 P_0$, the fourth level can be the $^3 D_1$ state. With parameters $(\Delta,\Omega,\Omega_c,\Delta_c)=(2\pi)(10,1,40,70)$~MHz, the Rabi frequency of atoms with the addressing beam is two times larger than other sites. For a frequency drift $\delta_c/(2\pi)$ across 10 lattice sites from $-2.5$~MHz to $2.5$~MHz~\cite{bernien2017probing}, the induced infidelity is on the order of $0.2\%$. Such infidelity can be mitigated by calibrating the addressing beam's intensity with spatial light modulators (SLM) and digital micromirror devices (DMD)~\cite{zhang_scaled_2023}.

\begin{figure}[htbp]
\includegraphics[width=0.49\textwidth]{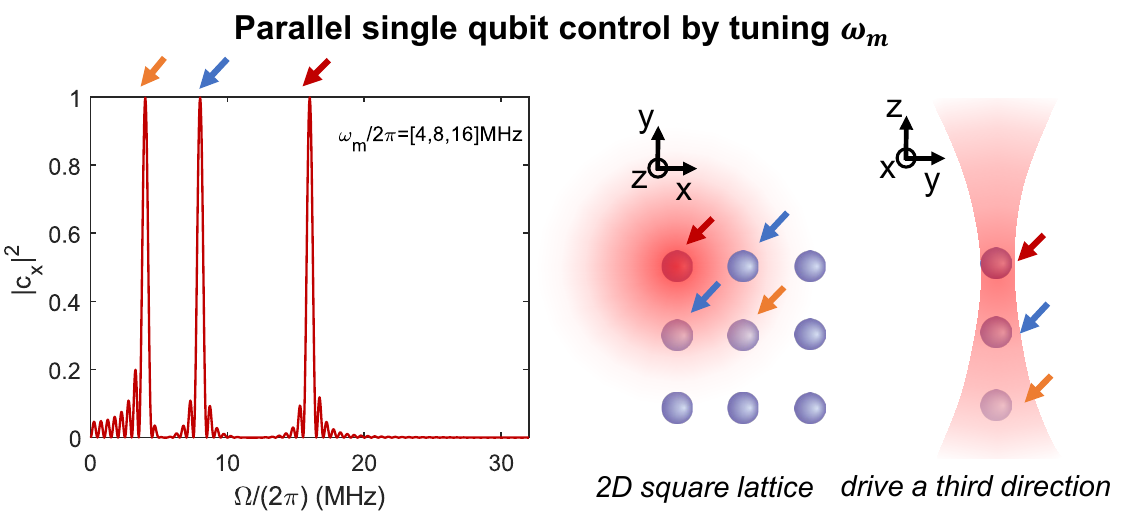}
\caption{\label{Fig3} \textbf{Parallel control of atom array with multiple modulation frequencies.} In the simulation we set $\epsilon_m/(2\pi)=0.125~$MHz, the maximum driving strength of the laser is $\Omega/(2\pi)=16~$MHz.}
\end{figure}

\subsection{Parallel control by adding multiple modulations}
With a broad laser beam focused on a small group of atoms, we selectively drive the target qubits at different sites by tuning the modulation frequency $\omega_m$ to the corresponding Rabi frequency $\omega_m=\Omega(r)$, which in turn, are controlled directly by laser intensity for optical qubits, or additional light shifts for hyperfine qubits. Actually, such an operation can be engineered in a highly efficient and parallel manner. Instead of sequentially applying each PM gate with evolution described by Eq.~\eqref{HI_PhaseMod}, we can engineer phase-modulated driving with multiple modulation frequencies described by $\Omega(r)\cos\left(\omega t +\phi-\sum_{i=1}^n{2\epsilon_m^{(i)}}/{\omega_m^{(i)}}\sin(\omega_m^{(i)} t+\phi_m^{(i)})\right)$ such that the Hamiltonian in the rotating frame is
\begin{equation}
    \ham_I^{(par)}=\frac{\Omega}{2}\sigma_x^\prime+\sum_{i=1}^n\epsilon_m^{(i)}\cos(\omega_m^{(i)}t+\phi_m^{(i)})\sz.
    \label{HI_PhaseMod_par}
\end{equation}
The modulation frequency $\omega_m^{(i)}$ is set to match the driving amplitude at site $\Omega(r^{(i)})=\omega_m^{(i)}$ for all target sites $(i)$ where parallel gates are desired. The modulation frequencies need to satisfy the condition $\omega_m^{(i)}=(2\pi k^{(i)}/\varphi^{(i)})\epsilon_m^{(i)}$ with $k^{(i)}$ integer numbers to engineering a $\varphi$ rotation for sites $(i)$. For example, to engineer $\pi$ rotations for different sites with the same total phase modulation time, $\epsilon_m^{(i)}=\epsilon_m=\pi/T$ for any $i$ and should be chosen as the common dividers of different $\Omega(r^{(i)})$, which slows down the gate speed. We note that such a slowdown is partially made up by the parallelism of the method, and the scenarios where atoms are arranged in a special lattice structure. In Fig.~\ref{Fig3}, we use numerical simulations to demonstrate the feasibility of such a parallel control technique applied to a small group of atoms. To control a larger group of atoms in parallel, one can move the beam center to a low-symmetry spatial point such that all sites have different and distinguishable driving amplitudes. To selectively control a subset of atoms in the array, one can switch on their corresponding phase modulation frequencies (matching their driving amplitudes, respectively).

The phase modulation scheme we develop is also useful to control a 3D atom array when crosstalk effects in the third ($z$) dimension are even larger (or unavoidable) than in the two-dimensional case as shown in Fig.~\ref{Fig3}. Moreover, phase modulation devices usually have much faster switching speeds than SLMs, showing advantages in achieving fast control in large-scale atom array quantum platforms.

\section{Avenues for performance improvement}
\begin{figure}[htbp]
\includegraphics[width=0.49\textwidth]{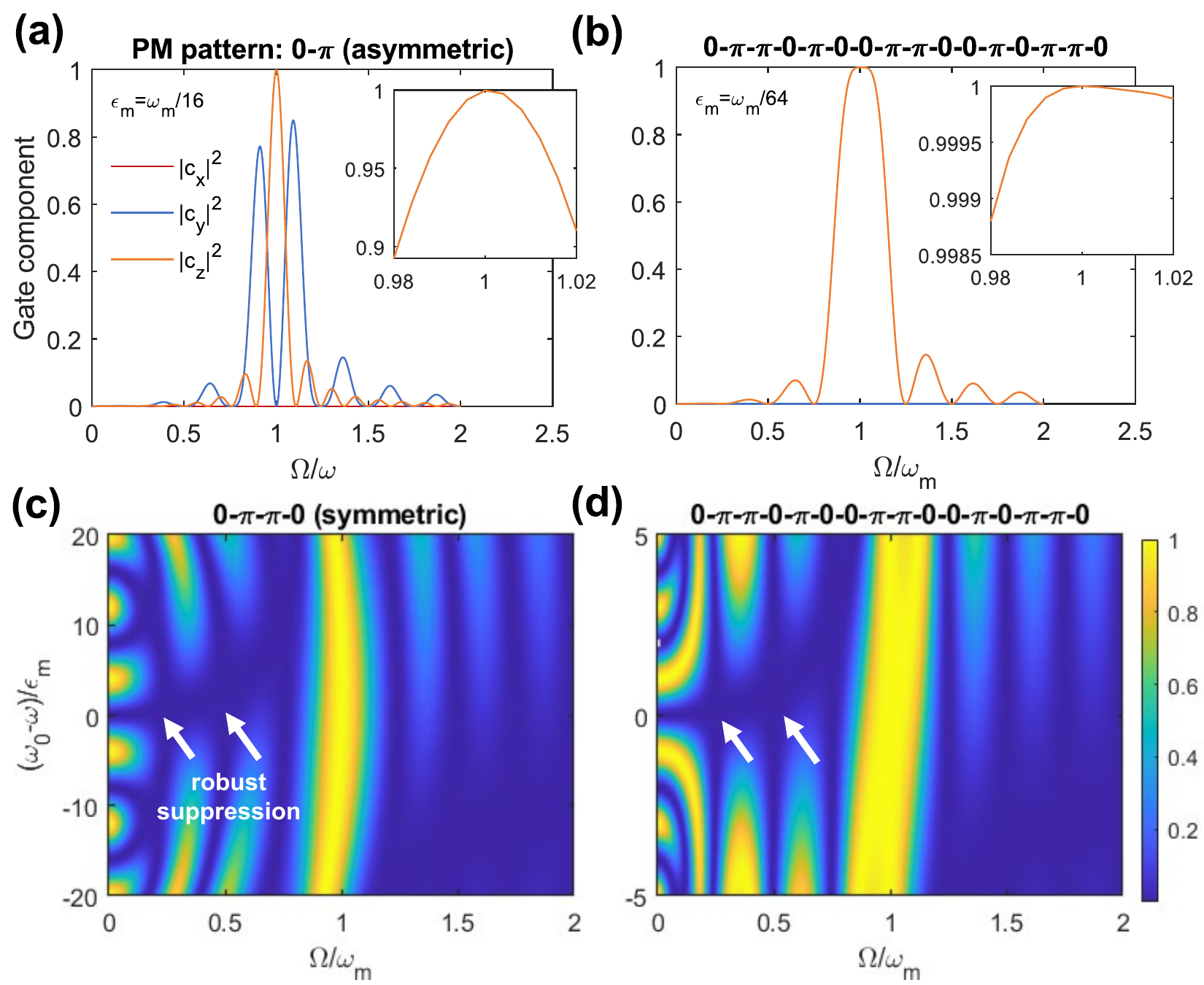}
\caption{\label{Fig4} \textbf{Gate improvement through sequence concatenation.} (a) The components of synthesized $Z$ gate. The phase for the first and second halves are set to $\phi=0,\pi$. (b) $Z$ gate with a phase pattern $0,\pi,\pi,0,\pi,0,0,\pi,\pi,0,0,\pi,0,\pi,\pi,0$ for 16 equal duration blocks. (c,d) 2D map of $\sz$ component as a function of both drive amplitude $\Omega$ and detuning $\omega_0-\omega$.}
\end{figure}

The performance of the crosstalk suppression can be further improved by concatenating the sequence to higher order. For the asymmetric $Z$ gate design in Fig.~\ref{Fig1}(a), the unitary evolution in the second rotating frame is 
\begin{align}
U^{(1)}=&[\cos^2\frac{\theta_t}{2}-\sin^2\frac{\theta_t}{2}\sigma_2\cdot\sigma_1] I +i\sin^2\frac{\theta_t}{2}[\sigma_1\times\sigma_2] \nonumber\\&+i\sin\theta_t\frac{\sigma_1+\sigma_2}{2}
\end{align}
where $\theta_t=\epsilon_R t$ is the rotation angle for each half period $t$ with a rate $\epsilon_R=\sqrt{(\Omega-\omega_m)^2+\epsilon_m^2}$, and $\sigma_{1(2)}=(\mp(\Omega-\omega_m)\sx^\prime-\epsilon_m\sz)/\epsilon_R$. 
We want to achieve an operation $U=-i\sz$ for $\Omega=\omega_m$  ($\sigma_{1,2}=\sigma_z$) and $U\approx I$ for $\Omega,\epsilon_m\ll\omega_m$, such that the crosstalk is suppressed. 
We note that similar bang-bang sequences have been the object of extensive work in optimal control~\cite{boscain_time_2006,aiello_algebraic_2015,aiello_qubit_2014,billig_optimal_2019,billig_time-optimal_2013}. 
Our goal is to cancel unwanted terms with sequence concatenation. Since the term $\propto \sigma_1\times\sigma_2$ changes sign with $\sigma_x^\prime$, 
it can be canceled by concatenating the sequence with the $0,\pi$ phase alternation. We denote $\Bar{U}^{(1)}$ an evolution similar to $U^{(1)}$ except for a $\pi$ phase shift in $\phi$ (and $\phi_m$ for $X$ gate design) such that the $\sx^\prime$ term in the Hamiltonian in Eq.~\eqref{eq:HI2} switches to $-\sx^\prime$. The symmetric $Z$ gate shown in Fig.~\ref{Fig1}(a) is actually constructed from the asymmetric one, through unitary $U^{(2)}=\Bar{U}^{(1)}U^{(1)}$. Higher order concatenation can be constructed in the same manner.
In Fig.~\ref{Fig4}, we show the gate improvement through sequence concatenation. As the concatenation order increases, the $\sy$ or $\sx$ components in the designed gate become smaller (Fig.~\ref{Fig4}(b)). Meanwhile, the $\sz$ curve gets more flattened in the center, making the target qubits immune to small driving amplitude fluctuations.

Our gate design is also robust against fluctuations of the frequency detuning. In Figs.~\ref{Fig4}(c,d) we show the Z gate fidelity (indicated by the intensity) as a function of both driving amplitude and frequency detuning, where the robustness of our design against both parameters is indicated by the broad regions pointed out by the white arrows.

To suppress crosstalk for a broad range of leakage power, a smaller driving strength $\epsilon_m$ is needed, which then makes the gate slower. Such a problem is usually unavoidable in narrowband pulse design since longer evolution is needed to narrow down the ``passband".  This drawback is partially made up by the enhanced coherence time under the phase-modulated driving, which makes the qubit insensitive to the dominant low-frequency noise~\cite{wangCoherenceProtectionDecay2020}. 
Moreover, when only the crosstalk on discrete sites needs to be suppressed (e.g., arrays in a lattice structure), we can make use of zero-crossing nodes and avoid choosing small $\epsilon_m$. Although our analysis is based on RWA, an effective Hamiltonian based on Floquet theory can be used to design the phase pattern analytically with high accuracy beyond the RWA~\cite{wang_sensing_2022,leskesFloquetTheorySolidstate2010,bartels_smooth_2013}. To efficiently optimize for discrete or continuous sequence parameters, one could also use numerical optimization methods or machine learning~\cite{low_methodology_2016,TianQuantumOptimalControlPhysRevA.102.043707,sauvage_optimal_2022}.

We now estimate the fidelity and crosstalk suppression performance of our gate design in a practical scenario with a control beam radius $r_0=7\mu$m and Rabi frequency $\Omega_0=(2\pi)2$MHz~\cite{levine_dispersive_2022}. If we consider a single atom with a temperature of 10~$\mu$K, the maximum infidelity is 0.004 and can be decreased to 0.001 with the first-order phase concatenation.  If we consider a worse scenario where a thermal atomic cloud with a radius of 1~$\mu$m leads to a Rabi frequency variation of about $1-e^{-1/7^2}=0.02$, the maximum gate infidelity is 0.1 and can be decreased to below 0.02 by the first-order concatenation design and below 0.001 by the third-order sequence concatenation. For both cases, the crosstalk effect in the nearest neighbor site for a square lattice is suppressed by more than two orders of magnitude. A more detailed calculation is included in Appendix~\ref{App:Fidelity}.

\section{Discussions} We developed a phase-modulated control technique to design single qubit gates that are robust against crosstalk effects due to the leakage of driving power. Strategies including sequence concatenation and hybrid gate design are developed to optimize the robustness against power and frequency fluctuations for the target qubits. Based on the proposed techniques, we use atom arrays as our target systems to develop two parallel single-qubit control schemes by either introducing additional addressing beams, or adding multiple modulation frequencies. Numerical simulations based on experimental parameters are added to show the possibility of implementing our schemes in current platforms. 

The typical narrow-band gate design based on composite pulses~\cite{torosov_smooth_2011,torosov_composite_2015,vitanov_arbitrarily_2011} could also be used for the control scheme with additional addressing beams. In comparison, the continuous modulated drive used in our protocol is compatible with the current electro-optic modulator devices, and the response to the driving amplitude in our design features discrete zero-crossing points at $3/4,1/2,1/4,\cdots$, thus are compatible with  atom arrays with a fixed geometry (e.g. square lattice). Moreover, the capability of adding multiple modulation frequencies in our design enables parallel gate operations and partially makes up for the slowdown of the gate speed, thus outperforming the composite pulse design only allowing sequential operation (see Appendix.~\ref{App:gate design}). Our results pave the way to controlling a 3D atom array~\cite{barredo_synthetic_2018}, which remains difficult due to the challenging optical paths design to avoid crosstalk. The phase-modulated CCD protects the coherence of Rabi oscillation against low-frequency noise~\cite{wangCoherenceProtectionDecay2020} and provides benefits in gate fidelity, which is critical in dynamical decoupling applications requiring a large number of single-qubit gates. Further improvement of the gate speed, fidelity, and phase smoothness by introducing more frequency modulations or machine learning is of interest for future research~\cite{TianQuantumOptimalControlPhysRevA.102.043707}.  Although this work focuses on single-qubit gate design for atom arrays, it provides insights in designing two-qubit gate without requiring the slow transport of atoms~\cite{bluvstein_quantum_2022}.

The presented technology might be generalized into solid-state quantum platforms such as superconducting and semiconducting qubits. In particular, control crosstalk in superconducting (SC) qubits~\cite{FluxCrosstalkPRApp2019,RigettiPRB2010,FluxCrosstalkPRQ2021,FluxoniumPRX2021} including crosstalk between dc-bias lines or microwave control lines has been one of the bottlenecks for scaling up SC-based quantum computers. For qubits whose frequencies are flux-controlled, the coupling from a bias line to an unintended loop and the targeted loop would induce non-negligible fluctuations in the qubit frequencies. This could then require cumbersome error calibration and mitigation processes~\cite{FluxCrosstalkPRQ2021}. The PM technique introduced here could be generalized into controlling the flux of qubits and would help reduce the Z-gate errors.

\acknowledgements
This work was in part supported by MIT-Harvard Center for Ultracold Atoms (NSF PHY 1734011).

\bibliography{main_text} 
\clearpage

\appendix
\section{Gate design with phase-modulated driving}
\label{App:gate design}
Here we provide further evidence of the performance of our strategies to achieve high-fidelity parallel gates.

\begin{figure}[htbp]
\includegraphics[width=0.495\textwidth]{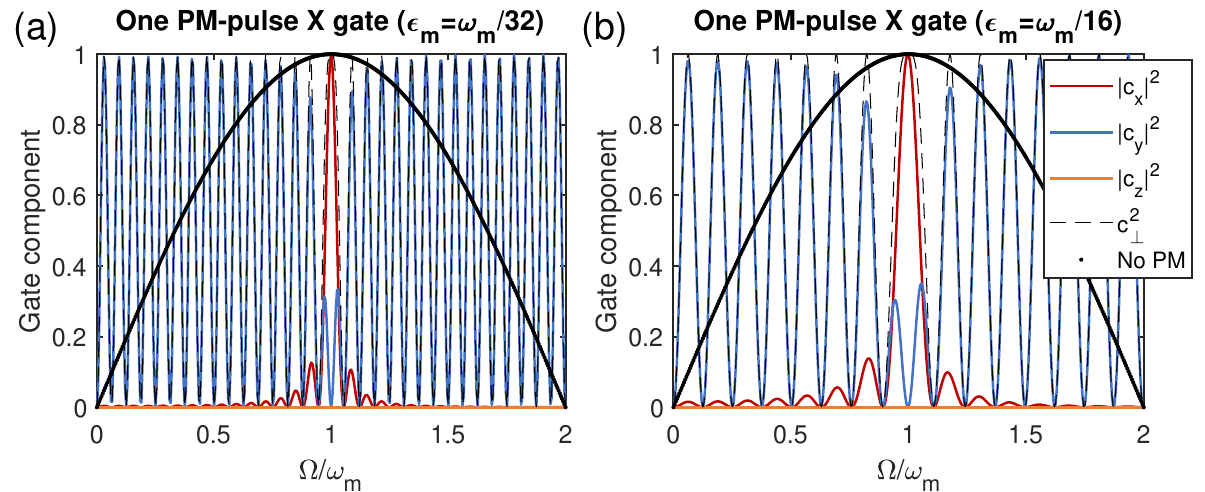}
\caption{\label{Fig2_2} Direct synthesis of the gate using one PM pulse. The phase is set to $\phi=\pi/2$ and the modulation phase is set to $\phi_m=\pi/2$. The duration of the pulse is set to $T=\pi/\epsilon_m$ and the value of $\epsilon_m=\omega_m/32,\omega_m/16$ for (a), (b) respectively.}
\end{figure}

In Fig.~\ref{Fig2_2}, we show the gate components of the $X$ gate engineered with a single pulse of phase-modulated drive without any refocusing. Although the $\sx$ component is suppressed when $\Omega\ll\omega_m$, a significant amount of $\sy$ component still exists.

\begin{figure}[htbp]
\includegraphics[width=0.495\textwidth]{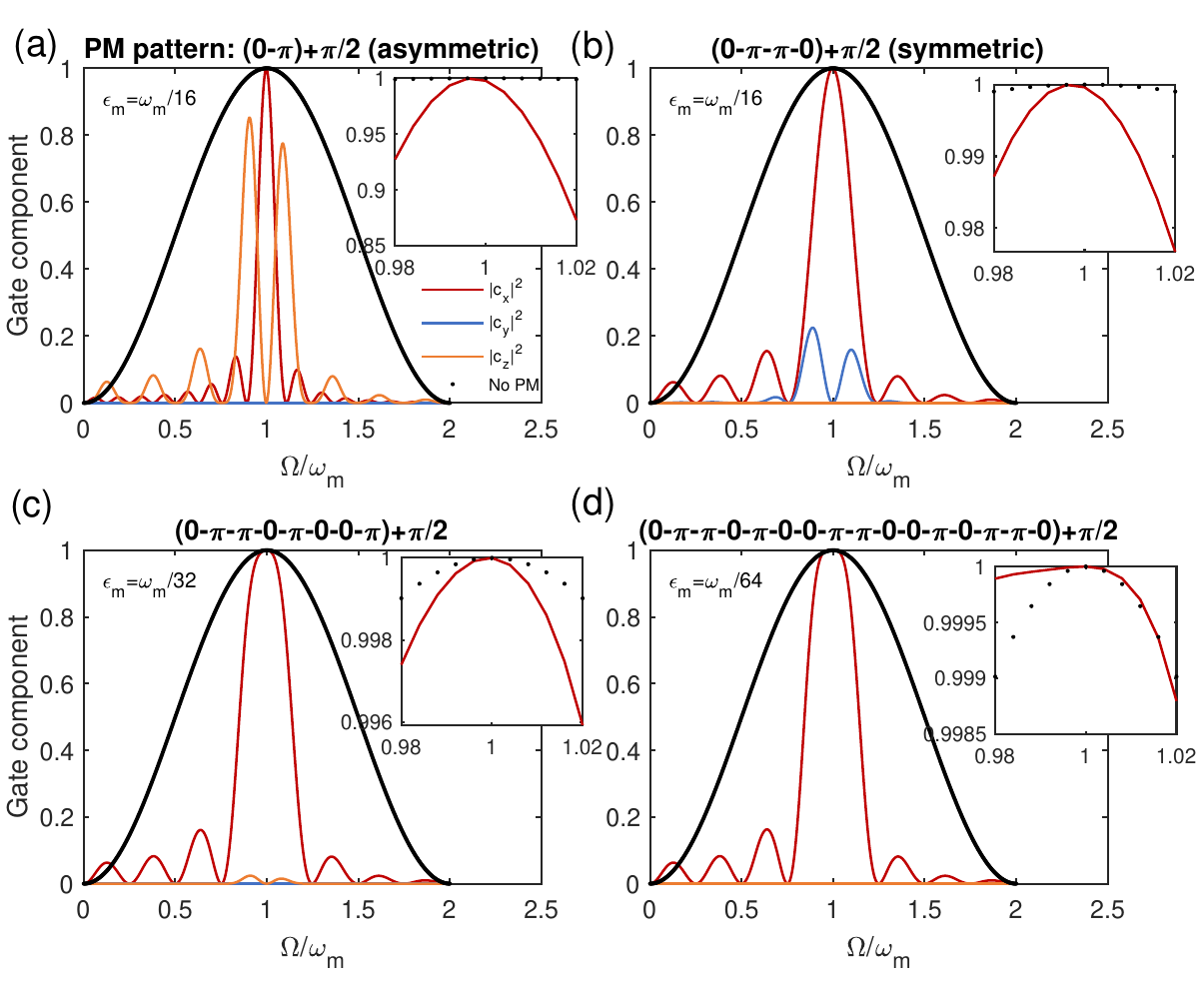}
\caption{\label{Fig2_3_X_PM} Comparison of different synthesized $X$ gate (PM design). (a) The phase for the first and second halves are set to $\phi=\pi/2,3\pi/2$ and $\phi_m=\pi/2,3\pi/2$, respectively. (b) Similar plot with (a) except for the $\phi,\phi_m$ phase pattern $\pi/2,3\pi/2,3\pi/2,\pi/2$ for four equal duration blocks. (c) The $\phi,\phi_m$ phase pattern is $\pi/2,3\pi/2,3\pi/2,\pi/2,3\pi/2,\pi/2,\pi/2,3\pi/2$ for eight equal duration blocks. (d) The $\phi,\phi_m$ phase pattern is $\pi/2,3\pi/2,3\pi/2,\pi/2,3\pi/2,\pi/2,\pi/2,3\pi/2$, $3\pi/2,\pi/2,\pi/2,3\pi/2,\pi/2,3\pi/2,3\pi/2,\pi/2$ for sixteen equal duration blocks.}
\end{figure}

\begin{figure}[htbp]
\includegraphics[width=0.495\textwidth]{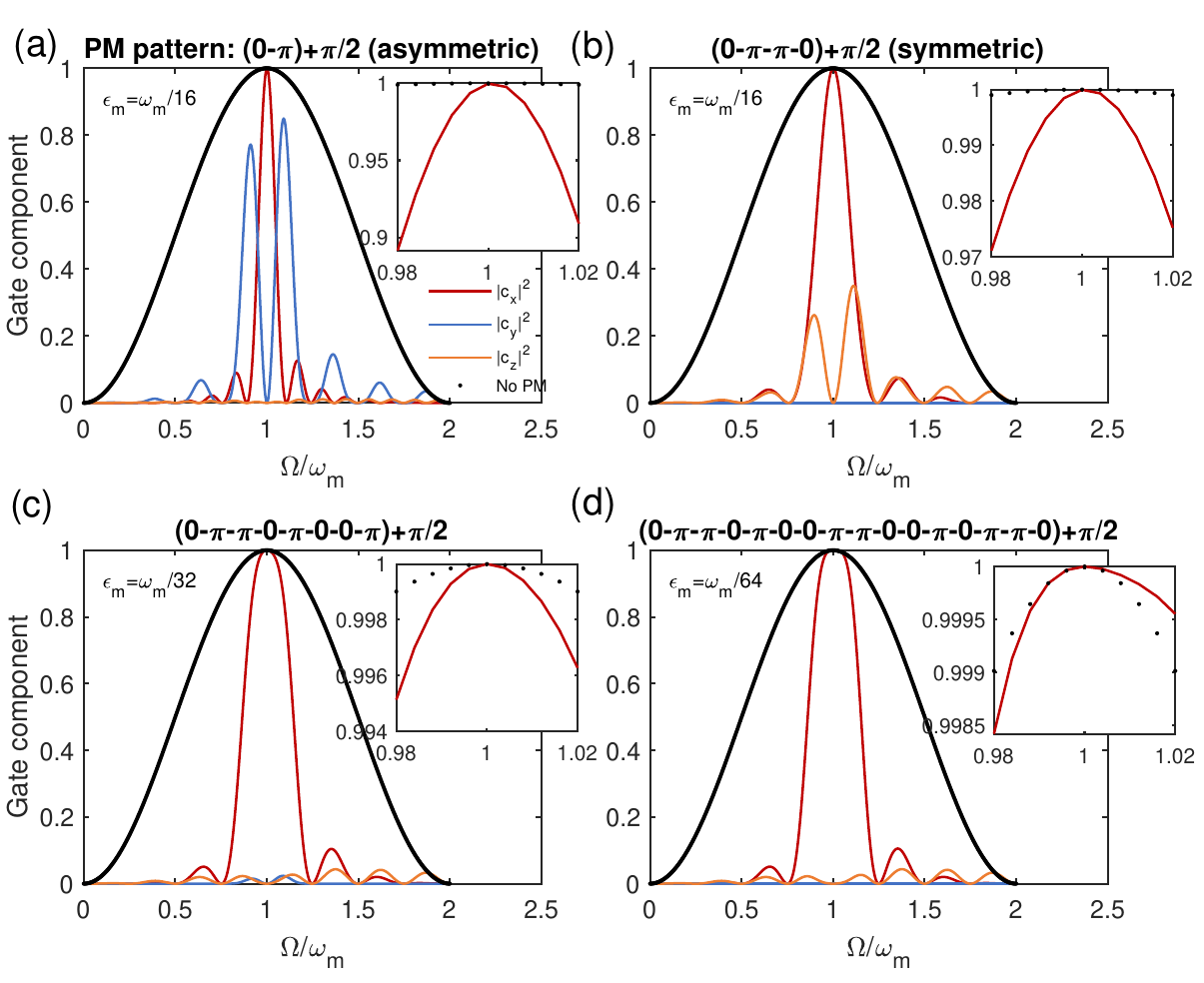}
\caption{\label{Fig2_3} Comparison of different synthesized $X$ gate (hybrid design). (a) The phase for the first and second halves of the PM $Z$ gate are set to $\phi=0,\pi$, with the modulation phase set to zero $\phi_m=0$ (same for b,c,d). (b) Similar plot with (a) except for the $\phi$ phase pattern $0,\pi,\pi,0$ for four equal duration blocks. (c) The $\phi$ phase pattern is $0,\pi,\pi,0,\pi,0,0,\pi$ for eight equal duration blocks. (d) The $\phi$ phase pattern is $0,\pi,\pi,0,\pi,0,0,\pi,\pi,0,0,\pi,0,\pi,\pi,0$ for sixteen equal duration blocks.}
\end{figure}

\begin{figure}[htbp]
\includegraphics[width=0.49\textwidth]{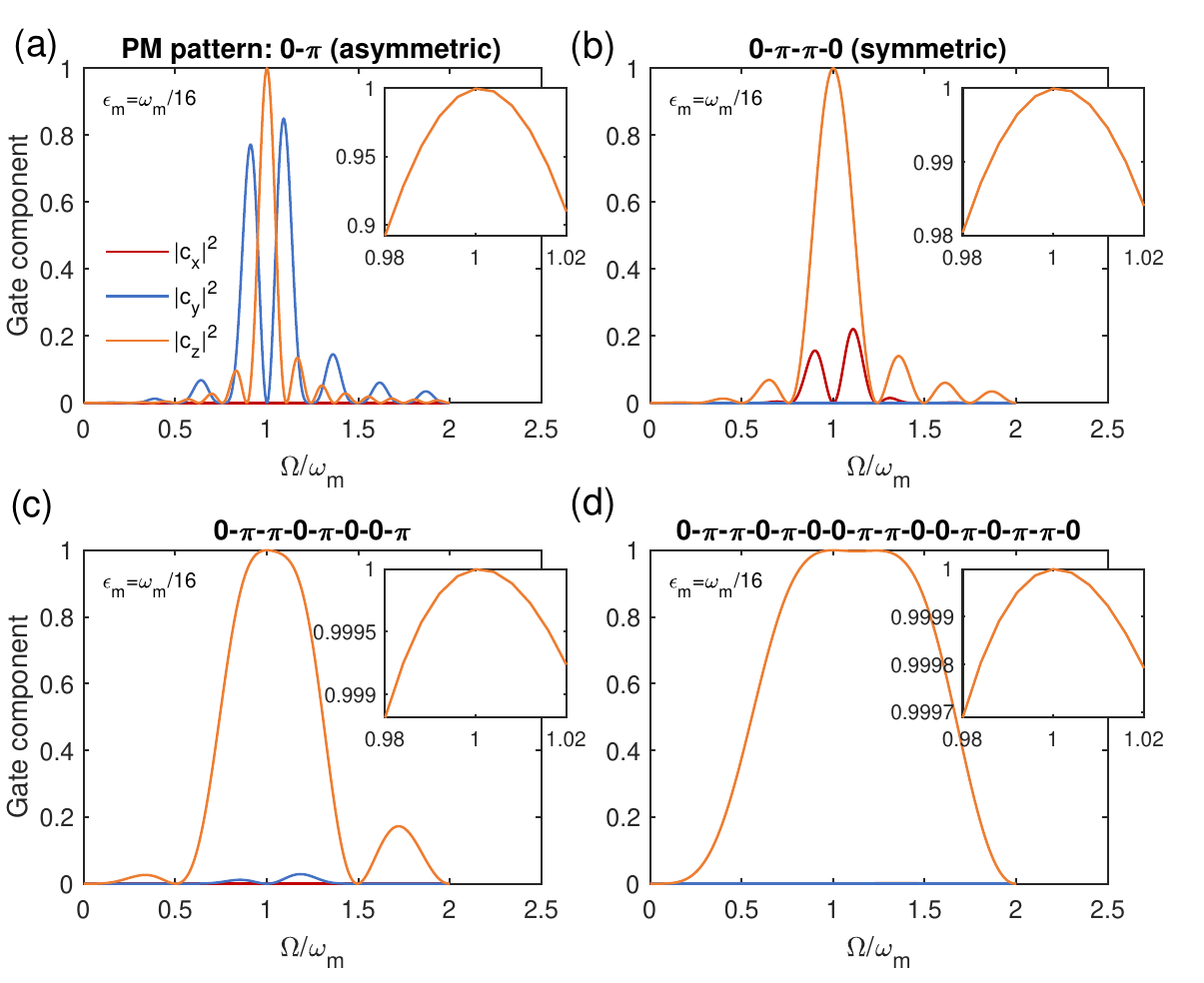}
\caption{\label{Fig2_3_ZGate_ep1o16} Comparison of different synthesized $Z$ gate under the fixed modulation strength $\epsilon_m=\omega_m/16$. }
\end{figure}

In Fig.~\ref{Fig2_3_X_PM}, we show the gate components of the designed $X$ gate composed of multiple pulses of PM drive under different concatenation orders. Although better performance can be obtained under higher concatenation order, the gate component for $\Omega\ll\omega_m$ is still non-negligible. In Fig.~\ref{Fig2_3}, we show the gate components of the designed $X$ gate based on the hybrid method (PM + normal pulses), which shows nice crosstalk suppression performance, especially under $\Omega\ll\omega_m$. When the order of the concatenation is increased, the unwanted $\sy,\sz$ components are better suppressed such that the $c_x$ component at $\Omega/\omega_m=1$ becomes  flatter. However, smaller $\epsilon_m$ is needed to maintain a similar profile. Such a tradeoff can be further seen in Fig.~\ref{Fig2_3_ZGate_ep1o16} where we fixed the $\epsilon_m$ and find a broader profile when we increase the concatenation order.  

\begin{figure}[htbp]
\includegraphics[width=0.49\textwidth]{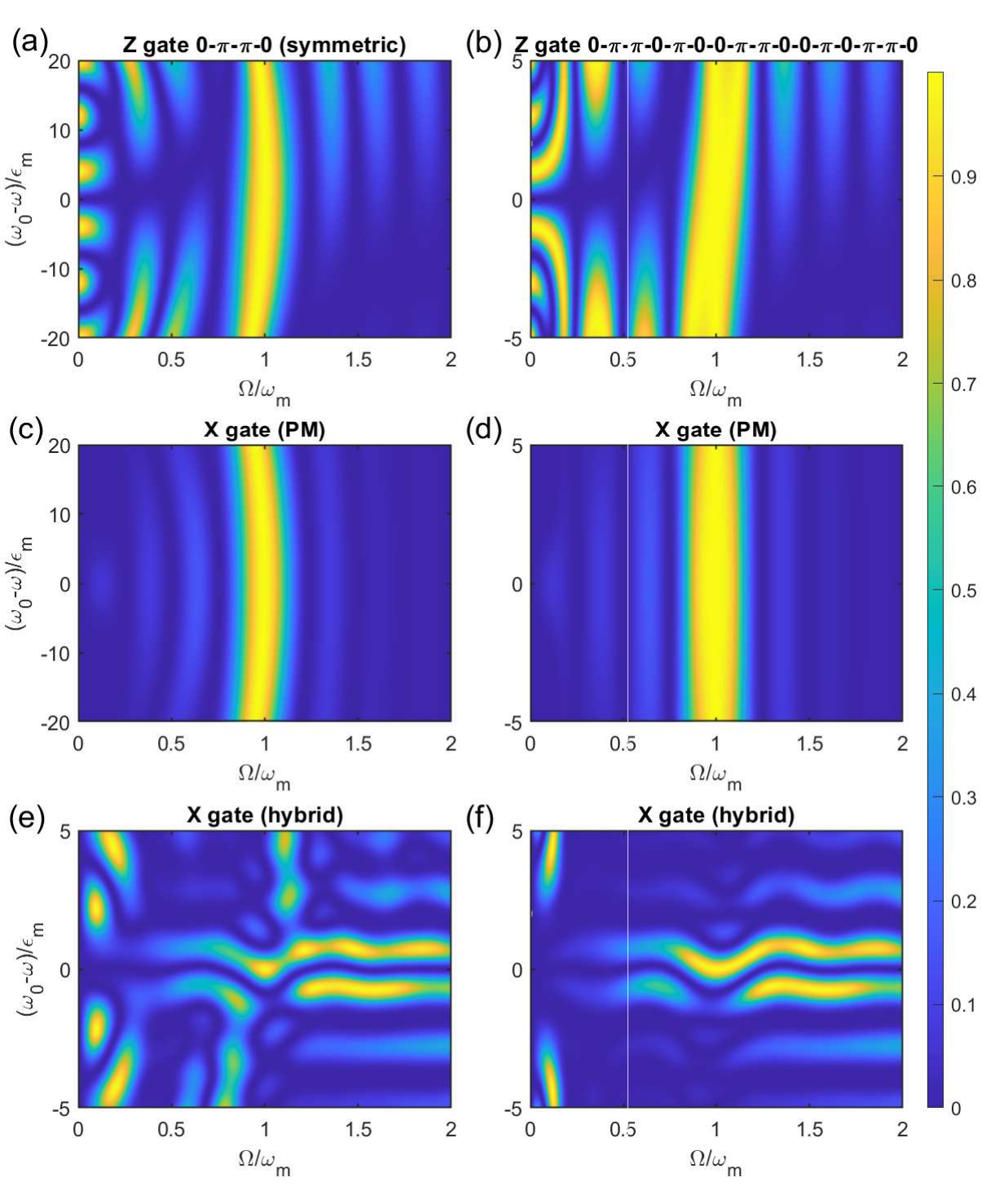}
\caption{\label{ZX_map} Comparison of different synthesized $Z$ and $X$ gates. For (a,c,e)  the modulation strength is set to $\epsilon_m/\omega_m=1/16$. For (b,d,f) the modulation strength is set to $\epsilon_m/\omega_m=1/64$.}
\end{figure}

\begin{figure}[htbp]
\includegraphics[width=0.49\textwidth]{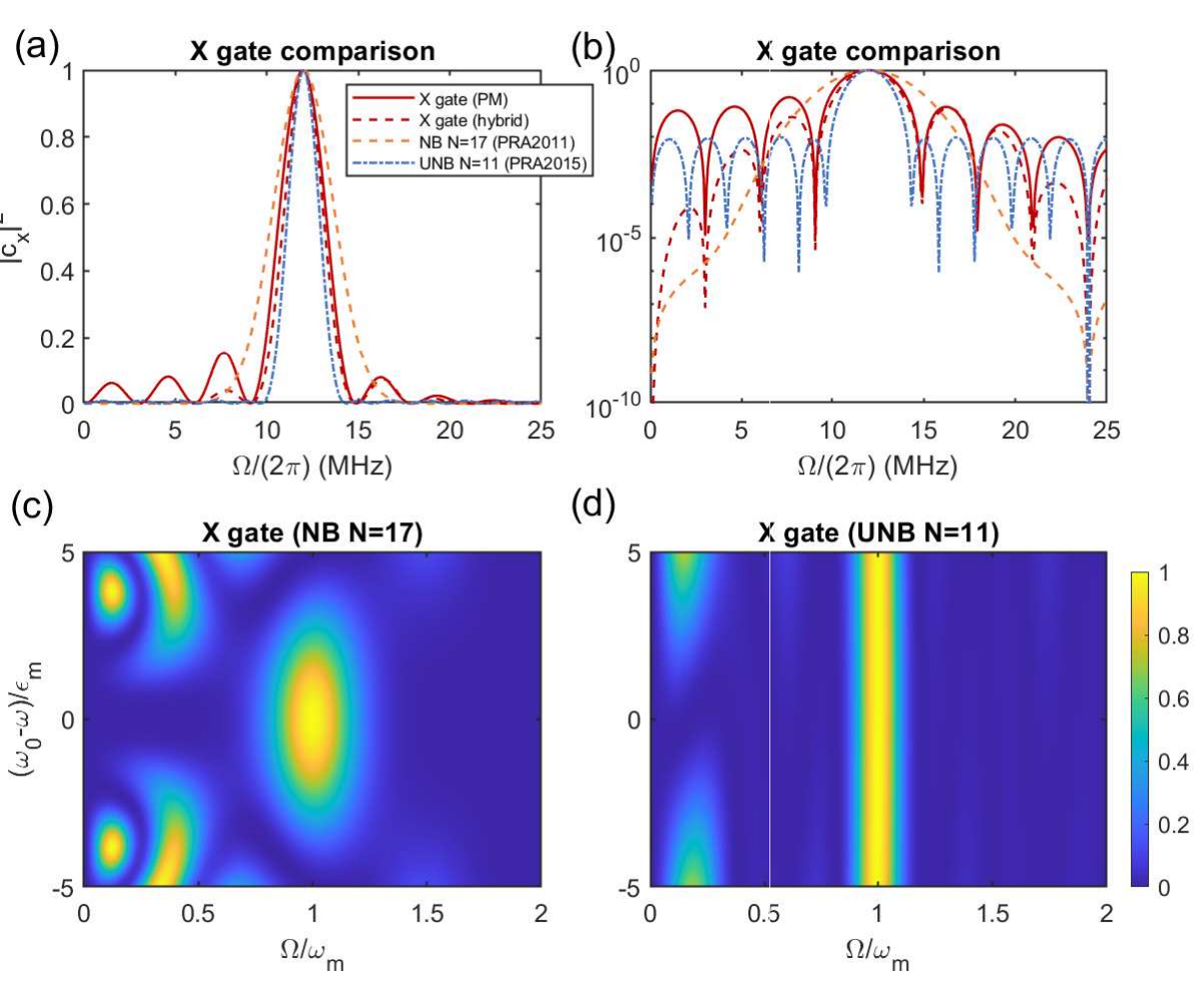}
\caption{\label{NBUNB} Comparison to narrowband composite pulses approach proposed in Refs.~\cite{torosov_composite_2015,torosov_smooth_2011,vitanov_arbitrarily_2011}. }
\end{figure}

We also include the simulation of different gate designs' response to both the change in driving amplitude $\Omega$ and detuning $\omega_0-\omega$, which are shown in Fig.~\ref{ZX_map}. We can see that clear dips are shown in all the gate design cases with different features. For the PM design of $Z$ gate and $X$ gate, the gate is not sensitive to the detuning and only sensitive to the amplitude change. While for the hybrid design of $X$ gate, the gate is both sensitive to the detuning and amplitude.

As a comparison to existing composite pulse methods, in Fig.~\ref{NBUNB} we plot our two $X$ gate designs together with the narrowband (NB) and ultra-narrowband (UNB) designs reported in Refs.~\cite{torosov_smooth_2011,torosov_composite_2015}. We use similar gate speeds (up to the same order of magnitude) for different methods. Different methods show quite different features, which could be used for different application scenarios. Our method provides nice crosstalk suppression at $\Omega/\omega_m=3/4,1/2,1/4,\cdots$ while sacrificing the smooth feature of the profile in comparison to the NB and UNB methods. We note that combining with numerical optimization and introducing more frequency terms in the phase modulation could provide better performance than our current development, which is out of the focus of this work.

\clearpage
\section{Parallel control with additional light shifts}
\label{App:light shifts}
Here we include details on the derivation of the parallel control with individual addressing control. We consider the four-level energy structure introduced in the main text denoted by $\ket{\uparrow},\ket{\downarrow},\ket{e},\ket{s}$ with energies $-\omega_1,\omega_2,0,\omega_3$. We apply circularly polarized drive field with frequencies $\omega_1+\Delta,\omega_2+\delta+\Delta,\omega_3+\Delta_c$ to couple the three levels $\ket{\uparrow},\ket{\downarrow},\ket{s}$ to the intermediate state $\ket{e}$. The Hamiltonian in the lab frame can be  written as
\begin{widetext}
\begin{equation}
H=
\begin{pmatrix}
-\omega_1 & \frac{1}{2} {\Omega_1}e^{i(\omega_1+\Delta)t} & 0 & 0\\
\frac{1}{2} {\Omega_1}e^{-i(\omega_1+\Delta)t} & 0 &\frac{1}{2} {\Omega_c}e^{i(\omega_3+\Delta_c)t} & \frac{1}{2} {\Omega_2}e^{-i(\omega_2+\delta+\Delta)t}\\
0 & \frac{1}{2} {\Omega_c}e^{-i(\omega_3+\Delta_c)t} & -\omega_3 & 0\\
0 & \frac{1}{2} {\Omega_2}e^{i(\omega_2+\delta+\Delta)t} & 0 & -\omega_2
\end{pmatrix}.
\end{equation}
\end{widetext}
In a rotating frame defined by $H_0=diag(-\omega_1,\Delta,\omega_3+\Delta_c+\Delta,-\omega_2-\delta)$, we obtain the Hamiltonian
\begin{equation}
H_I=
\begin{pmatrix}
0 & \Omega_1/2 & 0 & 0\\
\Omega_1/2 & -\Delta & \Omega_c/2 & \Omega_2/2\\
0 & \Omega_c/2 & -\Delta-\Delta_c & 0\\
0 & \Omega_2/2 & 0 & \delta
\end{pmatrix}.
\label{eq_supp:HI}
\end{equation}

Since the two intermediate states $\ket{e}$ and $\ket{s}$ are mixed with each other due to the additional addressing $\Omega_c$, we develop the following approach to calculate the effective Rabi rate between the two qubit states $\ket{\uparrow}$ and $\ket{\downarrow}$. We define a unitary rotational matrix $U_R$ to diagonalize the space spanned by $\ket{e},\ket{s}$, which gives rise to a new set of eigenbasis
\begin{align}
    \ket{\uparrow}&=\ket{\uparrow},\quad \ket{\downarrow}=\ket{\downarrow}\\
    \ket{+}&=\frac{(\Delta_c-\sqrt{\Delta_c^2+ \Omega_c^2})\ket{e}+\Omega_c\ket{s}}{ \sqrt{{(\Delta_c-\sqrt{\Delta_c^2+ \Omega_c^2})^2}+{\Omega_c^2}}}, \\
    \ket{-}&=\frac{(\Delta_c+\sqrt{\Delta_c^2+ \Omega_c^2})\ket{e}+\Omega_c\ket{s}}{ \sqrt{{(\Delta_c+\sqrt{\Delta_c^2+ \Omega_c^2})^2}+{\Omega_c^2}}}. 
\end{align}
such that the Hamiltonian is given by
\begin{equation}
H_I=U_RH_IU_R^\dagger=
\begin{pmatrix}
0 & \Omega_1^{+}/2 & \Omega_1^{-}/2 & 0\\
\Omega_1^{+}/2 & -\Delta^{+} & 0 & \Omega_2^{+}/2\\
\Omega_1^{-}/2 & 0 &  -\Delta^{-}  & \Omega_2^{-}/2\\
0 & \Omega_2^{+}/2 & \Omega_2^{-}/2 & \delta
\end{pmatrix}.
\end{equation}
The detunings and driving amplitudes in the new frame are then
\begin{align}
    \Delta^+&=\Delta+\frac{\Delta_c+\sqrt{\Omega_c^2+\Delta_c^2}}{2},\\
    \Delta^-&=\Delta+\frac{\Delta_c-\sqrt{\Omega_c^2+\Delta_c^2}}{2},\\
    \frac{\Omega_1^+}{\Omega_1}=\frac{\Omega_2^+}{\Omega_2}&=\frac{\Delta_c-\sqrt{\Delta_c^2+\Omega_c^2}}{\sqrt{(\Delta_c-\sqrt{\Delta_c^2+ \Omega_c^2})^2+\Omega_c^2}},\\
    \frac{\Omega_1^-}{\Omega_1}=\frac{\Omega_2^-}{\Omega_2}&=\frac{\Delta_c+\sqrt{\Delta_c^2+\Omega_c^2}}{\sqrt{(\Delta_c+\sqrt{\Delta_c^2+ \Omega_c^2})^2+\Omega_c^2}}.
\end{align}
The two-photon Rabi transition between $\ket{\uparrow}$ and $\ket{\downarrow}$ can be mediated by both of the states $\ket{+}$ and $\ket{-}$, while two transition paths are independent with each other. Thus, the effective Rabi rate is then the sum of the two paths, yielding
\begin{align}
\Omega_\text{eff}&=\frac{\Omega_1\Omega_2}{2}\left[(\frac{\Omega_1^+}{\Omega_1})^2\frac{1}{\Delta^+}+(\frac{\Omega_1^-}{\Omega_1})^2\frac{1}{\Delta^-}\right]\nonumber\\&=\frac{\Omega_1\Omega_2}{2}\times\frac{4(\Delta+\Delta_c)}{4\Delta(\Delta+\Delta_c)-\Omega_c^2}\nonumber\\&=\frac{\Omega_1\Omega_2}{2\Delta-\frac{\Omega_c^2}{2(\Delta+\Delta_c)}}.
\label{eq_supp:Omega_eff}
\end{align}

\begin{figure}[htbp]
\includegraphics[width=0.495\textwidth]{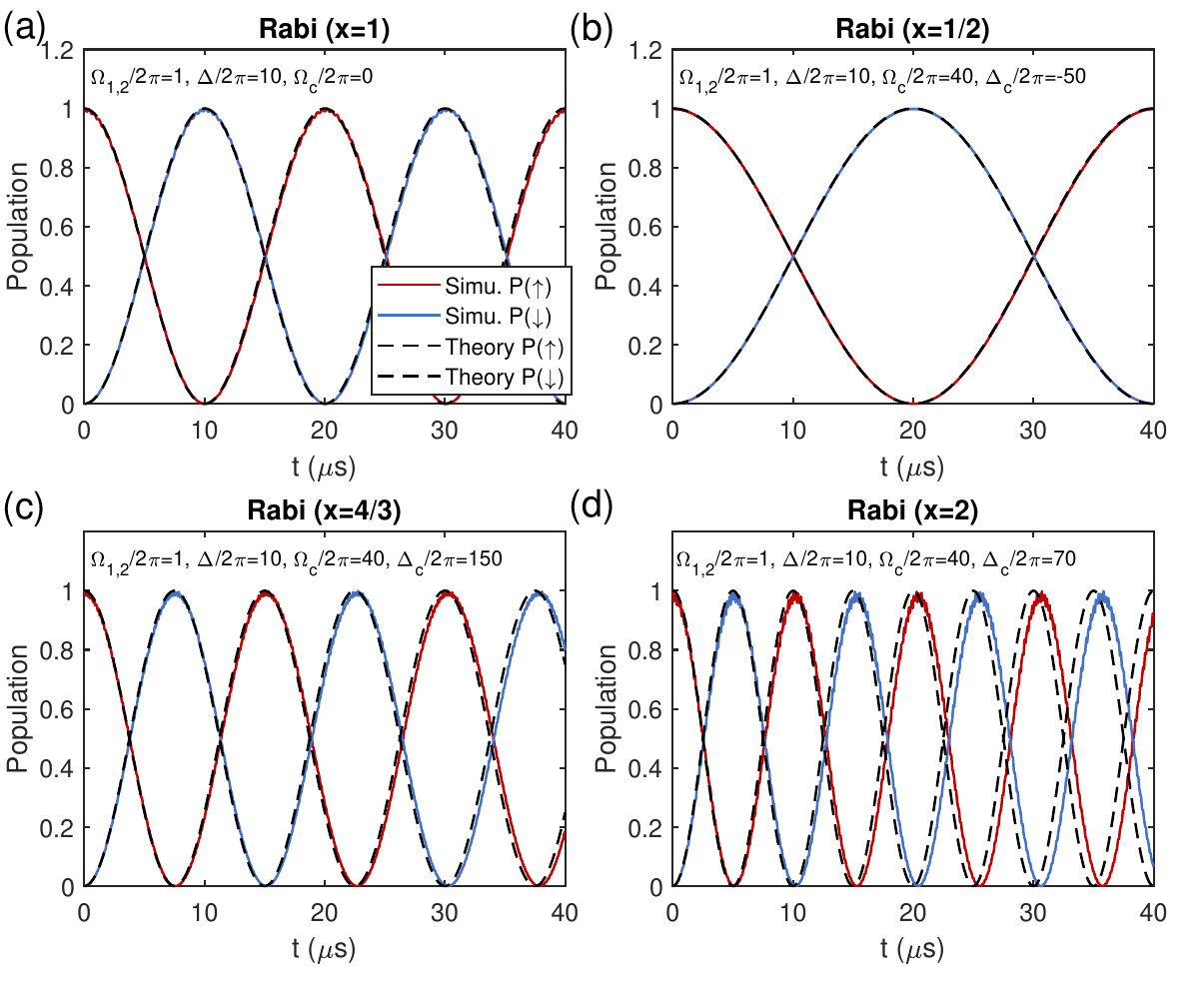}
\caption{\label{half} Tuning two-photon Rabi frequency with detuning $\Delta_c$. The simulation of the time-dependent population on $\ket{\uparrow}$ and $\ket{\downarrow}$ is calculated by directly evolving the system under the Hamiltonian in Eq.~\eqref{eq_supp:HI}. The theoretical prediction shown by the dashed lines are plotted using the effective Rabi rate in Eq.~\eqref{eq_supp:Omega_eff}.}
\end{figure}

If we require the effective Rabi for target qubits to be modified by a factor of $\xfactor$ in comparison to others without additional addressing $\Omega_c$ such that $1/(\Delta-\frac{\Omega_c^2}{4(\Delta+\Delta_c)})=\xfactor/\Delta$, we can obtain $\Delta_c=-\Delta+\frac{\xfactor\Omega_c^2}{4\Delta(\xfactor-1)}$. 
For example, to half the effective Rabi rate with the dressing beam, one can choose the detuning of the individual addressing beam as $\Delta_c=-\Delta-\frac{\Omega_c^2}{4\Delta}$; to double the effective Rabi rate, one can choose $\Delta_c=-\Delta+\frac{\Omega_c^2}{2\Delta}$. We implement numerical simulation to validate the effective Rabi rate we derived above. In Fig.~\ref{half}, we show four cases with $\xfactor=1,1/2,4/3,2$, where the numerical simulation matches the theoretical derivation. Though a very small inconsistency could exist due to the second-order perturbation approximation we used in deriving the two-photon Rabi rate, numerical simulation is always an efficient way to find out the parameters we need in experiments.

Now we analyzed  the sensitivity to the amplitude of the dressing beam, $\Omega_c$. To achieve arbitrary individual addressing, acoustic-optic devices are used, which correlates the positioning of beams to frequency shift. To first order, the relative change of the effective Rabi rate is $\frac{\partial \xfactor}{\partial \Delta_c}\delta_c=-\frac{4 \Delta (1-\xfactor)^2}{\Omega_c^2} \delta_c$, with $\delta_c$ the frequency difference towards the central beam. Therefore, small $\Delta$ and large $\Omega_c$ are preferred. Here we consider a realistic parameters with $^{174}$Yb atoms. The two hyperfine ground states of Yb are coupled through the triplet clock state $^3 P_0$. The individual addressing beam couples the transition between $^3 P_0$ and $^3 D_1$ state. With $\Delta = (2\pi)10$~MHz,  $\Omega = (2\pi)1$~MHz, $\Omega_c = (2\pi)40$~MHz, and $\Delta_c=(2\pi)70$~MHz, the modified Rabi rate with the addressing beam becomes twice  the Rabi rate without addressing beam. The sensitivity towards $\delta_c$ is suppressed by a factor of $\frac{\Delta}{\Omega_c^2}=1/160$. Here we consider a case similar to \cite{bernien2017probing}, in which the frequency drift across 10 lattice sites ranges from $\delta_c= -2.5$MHz to $\delta_c= 2.5$MHz, then the infidelity induced by the frequency change is on the order of $0.2\%$. This infidelity can be mitigated if we calibrate the addressing beam's intensity by adjusting the corresponding radio frequency power across the array, such that the modified Rabi rate with addressing beam is uniform. 

\section{Performance estimation in practical experiments}
\label{App:Fidelity}
{Based on the decomposition of the single qubit unitary evolution $U=c_0I+c_x\sx+c_y\sy+c_z\sz$, the fidelity of the target $Z$ ($X$) gate is $F=\Tr{|U^\dagger \sz|}^2/d^2=|c_z|^2$ ($|c_x|^2$) with dimension $d=2$. Thus the fidelity of the designed gate is given by the corresponding gate component coefficient.}

In general, to improve the performance of crosstalk suppression, one can choose a smaller $\epsilon_m$, which leads to a narrower passband near the resonance condition $\Omega=\omega_m$. However, the narrower passband degrades the gate fidelity when the driving amplitude $\Omega$ has effective variations due to either the instability of the laser, or the thermal motion or spatial distribution of atom (cloud). These variations usually depends on the control beam size, trapping frequency, and temperature. The influence due to such variations can be eliminated by improving the passband flatness through sequence concatenation as introduced in the main text, while a slower gate speed ($\epsilon_m$) is usually needed to ensure a similar narrow bandwidth.

Now we estimate a performance bound of our gate design in a practical scenario. We consider a Rabi spatial profile $\Omega(r)=\Omega_0 e^{-r^2/r_0^2}$ with a radius of $r_0=7\mu$m, corresponding to a 7$\mu$m (9.9$\mu$m) Gaussian beam waist of for single (two) photon Rabi control. The Rabi frequency is $\Omega_0=(2\pi)2$MHz~\cite{levine_dispersive_2022} and modulation strength is set to $\epsilon_m=\Omega_0/16$. A single atom cooled to a temperature of 10~$\mu$K has a speed of about 44~mm/s (for Rubidium) and travels about 0.2~$\mu$m within a 4~$\mu$s gate duration, leading to a Rabi amplitude variation of up to $1-e^{-0.2^2/7^2}=0.002$. The gate infidelity for the simplest design is upper bounded to 0.004 and can be decreased to 0.001 with the first-order concatenation.  For a worse scenario, we consider an atomic cloud with a radius of 1~$\mu$m cooled down to about 10~$\mu$K temperature. The spatial variation of atoms dominates the Rabi amplitude variation, which is about $1-e^{-1/7^2}=0.02$. The maximum gate infidelity for the simplest design is 0.1 and can be decreased to below 0.02 by the first-order concatenation design and below 0.001 by the third-order sequence concatenation  (see Fig.~5 and in the main text and Figs.~\ref{Fig2_3_X_PM},\ref{Fig2_3},\ref{Fig2_3_ZGate_ep1o16} in Appendix~\ref{App:gate design}). 

\begin{figure}[htbp]
\includegraphics[width=0.495\textwidth]{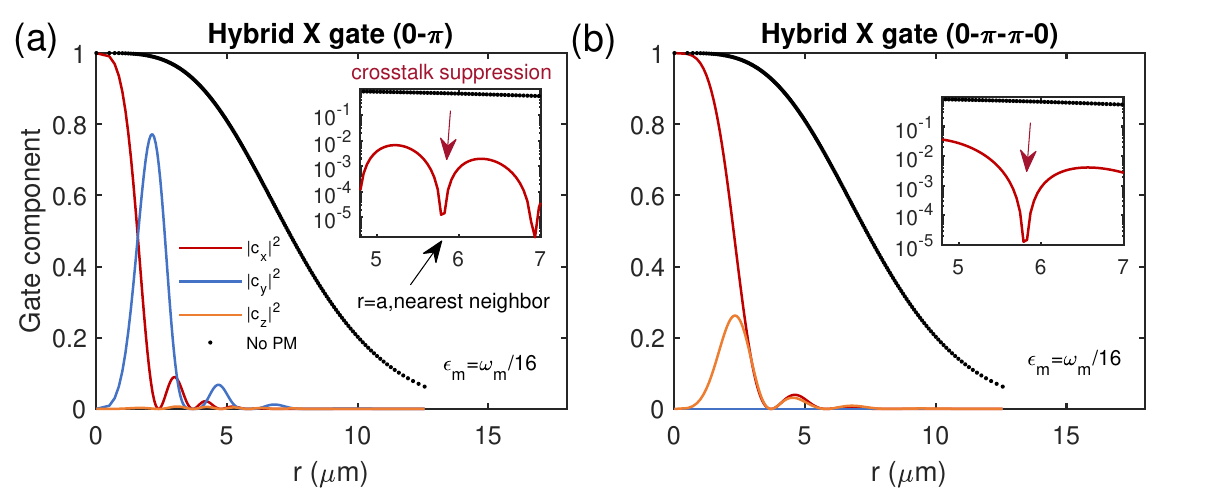}
\caption{\label{App_GateFidelity} Gate fidelity and crosstalk suppression effects in a practical setting assuming the radius of the Rabi amplitude profile is $r_0=7\mu$m. (a) Hybrid X gate design with the simplest phase pattern $0,\pi$. (b) Hybrid X gate design with the first-order concatenation of the phase pattern $0,\pi,\pi,0$. }
\end{figure}

To better visualize the crosstalk suppression performance at the nearest neighbor for a square lattice, we replot in Fig.~\ref{App_GateFidelity} the X gate performance shown by Figs.~\ref{Fig2_3}(a,b) as a function of the distance to the laser beam center $r$. The crosstalk effect in the nearest neighbor is suppressed by more than two orders of magnitude even when considering a 1$\mu$m spatial variation of atomic positions.

With infidelity of less than 0.001, appealing crosstalk suppression, and fast, parallel implementation capabilities, our gate design paves the way for building large-scale high-fidelity quantum computation platforms.

\end{document}